\documentclass[aps,prl,twocolumn,superscriptaddress]{revtex4}
\usepackage{graphicx}
\usepackage{supertabular}
\usepackage{color}
\usepackage{pdflscape}
\usepackage{gensymb}
\definecolor{red}{rgb}{0.75,0,0}
\definecolor{green}{rgb}{0,0.75,0}
\definecolor{blue}{rgb}{0.24,0.12,0.74}
\definecolor{darkblue}{rgb}{0.09,0.05,0.29}
\definecolor{notsodarkblue}{rgb}{0.18,0.10,0.58}
\definecolor{darkred}{rgb}{0.54,0.,0.}
\definecolor{grey}{rgb}{0.32,0.32,0.32}
\definecolor{lightgrey}{rgb}{0.44,0.44,0.44}

\usepackage{amsmath}
\usepackage{amsfonts}
\usepackage{amssymb}
\usepackage[title,titletoc,toc]{appendix}
\usepackage{natmove}
\usepackage{subfig}
\usepackage{graphicx}
\usepackage{caption}
\usepackage{booktabs}
\usepackage{floatpag}
\floatpagestyle{empty}

\usepackage{cmap}

\usepackage{import}
\usepackage[version=3]{mhchem} 
\usepackage{titlesec}

\newcommand{\scal}[2][]{{#2}_\mathrm{#1}}
\newcommand{\vect}[2][]{\boldsymbol{#2}_\mathrm{#1}}

\newcommand{\scali}[2][]{{#2}_{#1}}

\newcommand{\unit}[1]{\ensuremath{ \mathrm{#1} }}

\newcommand{\chem}[2][]{\ensuremath{ \mathrm{#2}_{#1} }}

\usepackage{lipsum} 
\usepackage[pdftex,dvipsnames]{xcolor} 
\usepackage[draft,prependcaption,textsize=tiny]{todonotes}

\usepackage{notes2bib}

\begin{document}


\title{Viscoelectric Effects in Nanochannel Electrokinetics}


\author{Wei-Lun Hsu}
\email{wlhsu@thml.t.u-tokyo.ac.jp}
\affiliation{Department of Mechanical Engineering, University of Tokyo, Tokyo 113-8656, Japan}
 \author{Dalton J. E. \surname{Harvie}}%
 \author{Malcolm R. Davidson}%
 \author{David E. Dunstan}%
  \affiliation{Department of Chemical and Biomolecular Engineering, University of Melbourne, Victoria 3010, Australia}%
 \author{Junho Hwang}%
 \author{Hirofumi Daiguji}%
\affiliation{Department of Mechanical Engineering, University of Tokyo, Tokyo 113-8656, Japan}



\begin{abstract}
Electrokinetic transport behavior in nanochannels is different to that in larger sized channels.  
Specifically, molecular dynamics (MD) simulations in nanochannels have demonstrated two little understood phenomena which are not observed in microchannels, being$\colon$ (i) the decrease of average electroosmotic mobility at high surface charge density, and (ii) the decrease of channel conductance at high salt concentrations, as the surface charge is increased.  However, current electric double layer models do not capture these results.  In this study we provide evidence that this inconsistency primarily arises from the neglect of the viscoelectric effect (being the increase of local viscosity near charged surfaces due to water molecule orientation) in conventional continuum models.  It is shown that predictions of electroosmotic mobility in a slit nanochannel, derived from a viscoelectric-modified continuum model, are in quantitative agreement with previous MD simulation results.  Furthermore, viscoelectric effects are found to dominate over ion steric and dielectric saturation effects in both electroosmotic and ion transport processes.  Finally, we indicate that mechanisms of the previous MD-observed phenomena can be well-explained by the viscoelectric theory.  

\end{abstract}


\maketitle

Electrokinetic transport of aqueous electrolytes in nanofluidic channels is of essential importance to a number of cutting-edge technologies such as capacitive deionizaion~\cite{Suss12},  nanofluidic batteries~\cite{Daiguji04} and bio-nanosensing~\cite{Dekker07}.
To gain a fundamental understanding of electrokinetics in these systems, both molecular dynamics (MD) and continuum theory have been conducted to investigate transport processes at the nanoscale~\cite{Shirono09, Daiguji10}.
As MD simulations directly utilize atomic properties when predicting transport behavior, one would expect these results to be more accurate than continuum based simulations, however high computational cost severely limits the dimensions of investigable systems. 
Within nanofluidic devices, although the channel size (\emph{i.e.}, height or diameter) is just a few nanometers, the length (and/or width) of these channels can be up to several microns or millimeters in many cases, which is computationally infeasible for MD~\cite{Kim10}. 
Besides, for many nanofluidic applications, nanochannels are integrated with microchannels (or reservoirs) that directly influence the transport behavior within the nanochannels, further expanding the required simulation domain~\cite{Hsu14}. 
Hence under most circumstances, continuum simulations must be relied on.

\begin{figure} 
\centering
\includegraphics[keepaspectratio, width=0.45\textwidth]{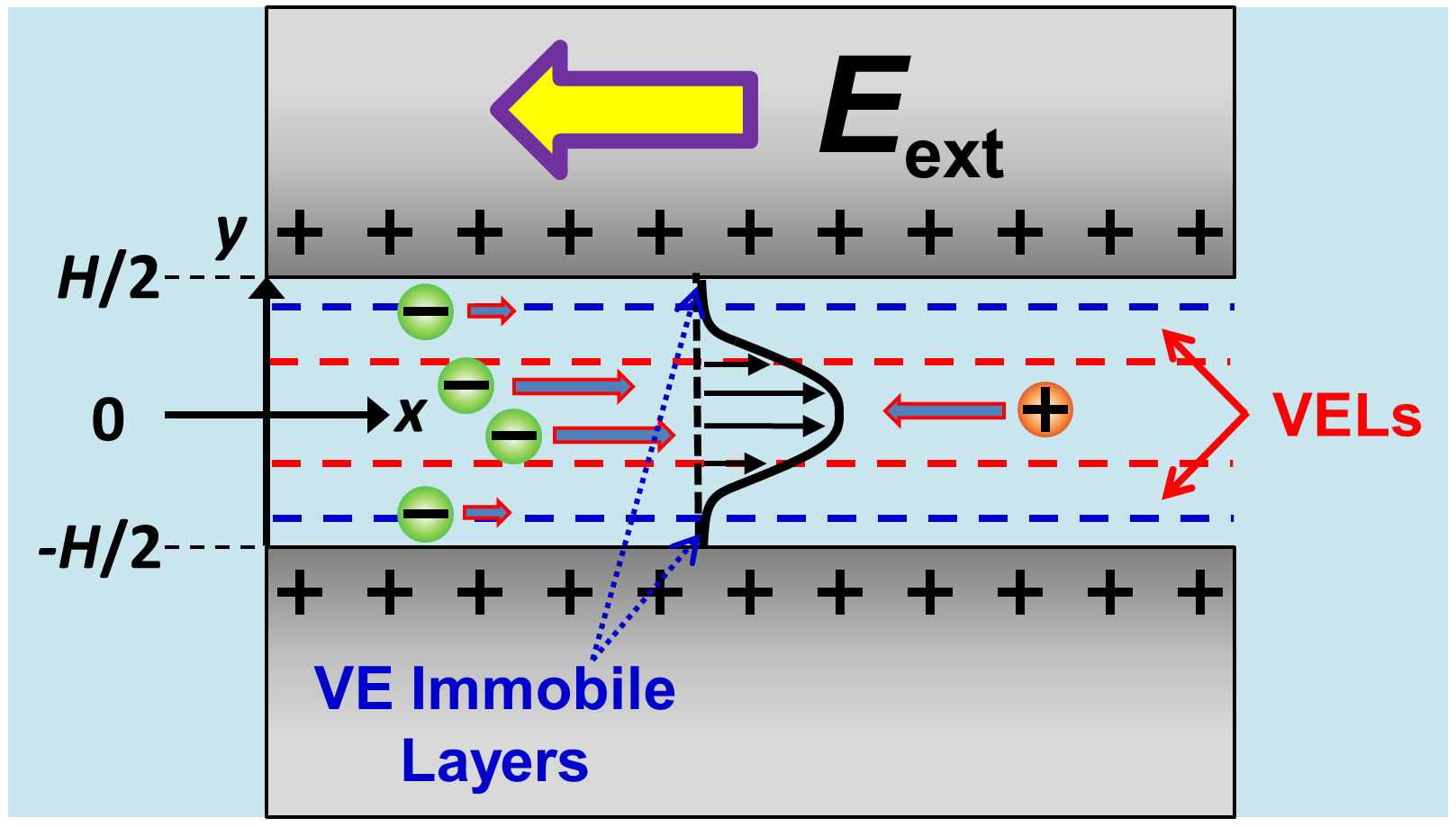}
\caption{\label{fig:wide} Schematic of the transport behavior in a positively charged slit nanochannel.
Within the viscoelectric layers (VELs), the viscosity is increased and ionic diffusivities and dielectric permittivity are decreased, attributed to the orientation of water molecules.
The solution is effectively immobile in the vicinity of the surface ($i.e.$ within the viscoelectric (VE) immobile layers) due to the high viscosity.
~$H$ denotes the channel height. }
\end{figure}

However, a lack of consistency between previous MD and continuum simulation results for nano-length nanochannels has been observed~\cite{Qiao03, Qiao05}, implying that the employed continuum models may be missing critical physical transport phenomena that are relevant in nanochannels. 
In particular, an increase in solvent viscosity and decrease in solute diffusivities (compared to their bulk values) in the vicinity of a charged surface are found in previous MD simulations~\cite{Freund02, Qiao05} (properties which are usually assumed constant in continuum models), resulting in an overestimated electroosmotic velocity and channel conductance by continuum theory.  
In addition, previous MD simulations~\cite{Qiao05} have reported two counter-intuitive transport phenomena in nanochannels$\colon$ Decreases in (i) electroosmotic mobility at high surface charge levels, and (ii) channel conductance at high salt concentrations, when increasing the surface charge. 
Detailed mechanisms of these unique phenomena have not been clarified previously, although it was suspected that they may relate to the increased viscosity~\cite{Wu11}.

To investigate full range nanofluidic systems with sufficient accuracy, and to better understand electrokinetic transport phenomena in nanochannels, a simple continuum model that captures the dominant behavior in MD simulations is desired but yet to be available. 
Thus, the objective of this study is to construct a modified continuum model that offers close predictions to previous MD results and, using it as a tool, we examine the mechanisms of the observed phenomena in water and ion transport.
The channel geometry and solution conditions considered are based on the previous MD study by~\citet{Qiao05} in which a potassium chloride (KCl) aqueous solution is confined in a long slit channel separated by a distance $H$, as illustrated in Figure~1.
Both surfaces possess the same amount of positive charge uniformly distributed along each wall.  
An external electric field~$\vect{E_\text{ext}}$ ($|\vect{E_\text{ext}}|$=~0.2~V/nm) is applied parallel to the surfaces, simultaneously yielding an electroosmotic flow in the opposite direction of~$\vect{E_\text{ext}}$ and an electric current in the same direction of~$\vect{E_\text{ext}}$, whereby the local nanochannel conductance in the $y$-direction per unit length along the channel $\scal[L]G(y)$ is obtained as$\colon$
\begin{equation}
\begin{aligned}
\scal[L]G(y) = e\left(\frac{v(y)}{\vect{|E_\text{ext}|}}+\frac{e\scal[K^+]{\mathcal{D}}}{\scal[B]{k}T}\right)\scal[K^+]{n}(y)\\
-e\left(\frac{v(y)}{\vect{|E_\text{ext}|}}-\frac{e\scal[Cl^-]{\mathcal{D}}}{\scal[B]{k}T}\right)\scal[Cl^-]{n}(y)
\end{aligned}
\end{equation}
in which $y$ is the direction normal to the surfaces with $y=0$ on the channel centerline, $e$ the element charge, $v(y)$ the electroosmotic velocity, $\scal[B]{k}$ the Boltzmann constant, $T$ the temperature (= 300~K), $\scal[K^+]{\mathcal{D}}$ the ionic diffusivity of $\chem{K^+}$,  $\scal[Cl^-]{\mathcal{D}}$ the ionic diffusivity of $\chem{Cl^-}$, $\scal[K^+]{n}(y)$ the $\chem{K^+}$  concentration and $\scal[Cl^-]{n}(y)$ the $\chem{Cl^-}$ concentration. 

We employ the electric Poisson equation and a modified Navier-Stokes equation (considering an electric body force from $\vect{E_\text{ext}}$) to calculate $\scal[K^+]{n}(y)$,  $\scal[Cl^-]{n}(y)$ and $v(y)\colon$    
    
\begin{equation}
\frac{d^2\phi(y)}{dy^2}  = - \frac{\rho_e}{\scal[r]\epsilon\scali[0]{\epsilon}} = - \frac{ \scal{z}e}{\scal[r]\epsilon\scali[0]{\epsilon}}\left(\scal[K^+]{n}(y)-\scal[Cl^-]{n}(y)\right)
\label{eq:Poisson1D}
\end{equation}

\begin{equation}
\frac{d}{dy}(\eta\frac{d{v(y)}}{dy}) + \rho_e|\vect{E_\text{ext}}| = 0
\label{eq:NS1D} 
\end{equation}
In these expressions, $\phi(y)$ is the electric potential, $\rho_e$ the space charge density, $\scal[r]\epsilon$ the relative permittivity of the solution, $\scali[0]{\epsilon}$ the permittivity of vacuum, $\eta$ the viscosity and $\scali{z}$ the ionic valence of binary electrolytes (=1 for the KCl solution).

The presence of ions affects the transport behavior in three ways$\colon$ 

(i)  Steric (S) effects$\colon$ 
Since the classical Boltzmann equation assumes the ions are point-like (which becomes invalid when the channel size is just several nanometers), a modified Boltzmann distribution considering steric effects of ions is employed and $\rho_e$ is expressed as~\cite{Borukhov97}$\colon$

\begin{equation}
 \rho_e = \frac{2ze\scal[0]{{n}}\sinh({\frac{\scali{z}e\phi}{\scal[B]{k}T}})}{1+2\lambda\sinh^2({\frac{\scali{z}e\phi}{2\scal[B]{k}T}})} 
\end{equation}
where $\scal[0]{{n}}$ is the bulk KCl concentration and the bulk volume fraction of ions $\lambda$ is given by$\colon$  
\begin{equation}
 \lambda = 2\scali[0]{{n}}a^3
\end{equation}
where $a$ is the size of hydrated ions and given by $a = 6.6~\AA$~\cite{Qiao05}.

(ii)  Dielectric (DE) effects$\colon$ 
Due to the high electric field near the surfaces, the water molecules are electrically saturated and thus the dielectric permittivity at the interface is lower than the bulk value~\cite{Debye29}.
A permittivity modification was proposed by Booth~\cite{Booth51} based on the Onsager~\cite{Onsager36} and Kirkwood~\cite{Kirkwood39} theories of polar dielectrics and re-interpreted by Hunter~\cite{Hunter66} as$\colon$
\begin{equation}
\scal[r]\epsilon = \scal[r,0]\epsilon(1-b|\vect{E_\text{EDL}}|^2)%
\end{equation}
where $\scal[r,0]\epsilon$ and $\vect{E_\text{EDL}}$ are the relative permittivity of the solution in the absence of an electric field (= 81) and local electric field within the electric double layer (EDL), respectively.  The coefficient $b$ is estimated to be 4$\times 10^{-18}~\unit{m^2/V^2}$ for water~\cite{Hunter66, Hunter81}.


(iii)  Viscoelectric (VE) effects$\colon$ 
As a result of the variation in vibration frequency of water molecules, the interactions of orientated water molecules near the charged surface increase.
Consequently, the motion of water molecules is largely inhibited, giving rise to a higher apparent viscosity.  
A formula was proposed by~Andrade and Dodd \cite{Andrade51} based on experimental observations at low electric field magnitude and, later theoretically verified by~\citet{Lyklema61} for water$\colon$
\begin{equation}
\eta = \eta_0(1+f|\vect{E_\text{EDL}}|^2)
\end{equation}
where $\eta_0$ and $f$ are the viscosity of the solution in the absence of an electric field (= 0.743 mPa$\cdot$s) and VE coefficient, respectively. 
Here, we extend eq~7 for arbitrary electric field magnitude based on the theory of polarization~\cite{Lyklema61}$\colon$
\begin{equation}
\begin{split}
\eta & = \eta_0\exp(\frac{\Delta E_\text{a}}{\scal[B]{k}T}) = \eta_0\exp(\frac{\alpha m^2{E^2_\text{i}}}{\scal[B]{k}T}) \\ 
&= \eta_0\exp(f|\vect{E_\text{EDL}}|^2) = \eta_0 \sum^\infty_{n=0}\frac{(f|\vect{E_\text{EDL}}|^2)^n}{n!}
\end{split}
\end{equation}
where $\Delta E_\text{a}$ is the increased activation energy due to the presence of $\vect{E_\text{EDL}}$, constant $\alpha$ a structural coefficient, $m$ the dipole moment, and $E_\text{i}$ the internal electric field magnitude which is proportional to $|\vect{E_\text{EDL}}|$.  
If $f|\vect{E_\text{EDL}}|^2$ $\ll$ 1, eq~8 converges to eq~7, however there is no theoretical basis for neglecting the higher order terms of eq 8 (generally for EDLs), and indeed, in this paper we find that their effect is significant (a typical value of surface $|\vect{E_\text{EDL}}|$ = 1.18~$\times 10^{8}~\unit{V/m}$, when the surface charge density equals 80 $\unit{mC/m^2}$, making $f|\vect{E_\text{EDL}}|^2 > 1$).
In this study, where we employ eq~8, it is found that $f$ = 2.3~$\times 10^{-16}~\unit{m^2/V^2}$  achieves the closest fit to MD simulation results, which is close to previous experimental estimates of $f$~\cite{note1}. 

Based on the Stokes-Einstein equation, the ionic diffusivity $\scali[i]{\mathcal{D}}$ (in which $i$ denotes $\chem{K^+}$ or $\chem{Cl^-}$) concurrently decreases due to the increase of hydrodynamic drag force on ions when migrating in a viscous solution~\cite{Hsu16}~$\colon$
\begin{equation}
\scali[i]{\mathcal{D}} = \frac{\scali[i,0]{\mathcal{D}} \eta_0}{\eta} 
\label{eq:SE}
\end{equation}
where $\scali[i,0]{\mathcal{D}}$ is the ionic diffusivity in the absence of an electric field (= $1.96~\times 10^{-9}~\unit{m^2/s}$ for $\chem{K^+}$ and $2.03~\times 10^{-9}~\unit{m^2/s}$ for $\chem{Cl^-}$). 
A decrease in diffusion coefficients of spherical nanoparticles within a quartz (negatively charged) nanopillar chip was experimentally observed by Kaji et al.~\cite{Kaji06} supporting eq 9.


On the boundaries, it is assumed that the nanochannel walls are non-conductive that the surface charge is entirely balanced by the net charge within the solution, and non-slip~\cite{Qiao05}. 
We derive the following boundary conditions at $y=\pm{H}/{2}\colon$ 
\begin{equation}
\frac{d\phi}{dy} = \mp\frac{\sigma}{\scal[r]\epsilon\epsilon_0}
\end{equation}
\begin{equation}
v = 0%
\end{equation}
in which $\sigma$ denotes the surface charge density. 


\begin{figure} [t]
\centering
\includegraphics[keepaspectratio, width=0.45\textwidth]{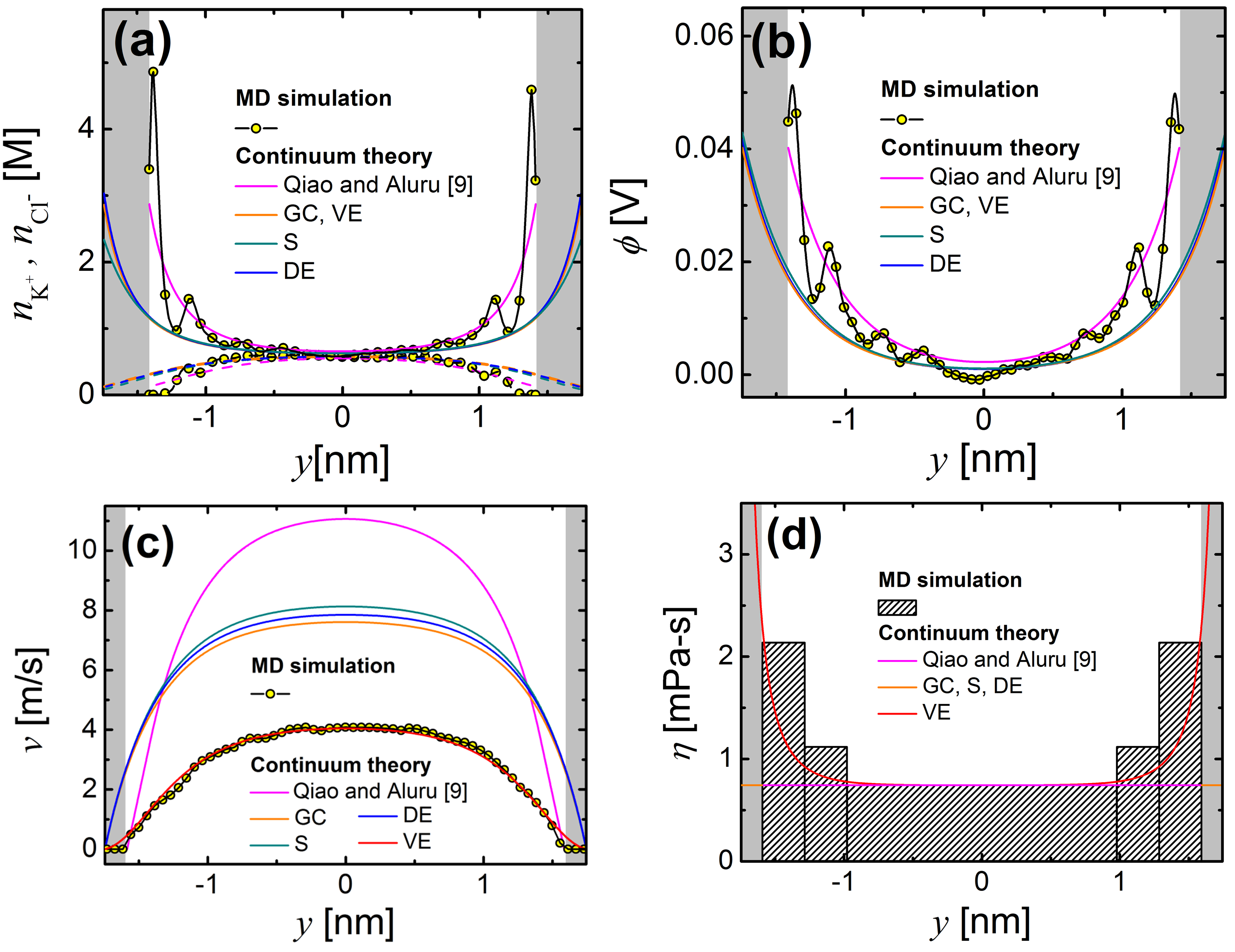} 
\caption{\label{fig:epsart}   (a) Ion concentrations $\scal[Cl^-]{n}$ (solid curves) and $\scal[K^+]{n}$ (dashed curves), (b) electric potential $\phi$~\cite{note2}, 
(c) electroosmotic velocity $v$ and (d) viscosity $\eta$ profiles along the $y$-direction at $\sigma$ = 80 $\unit{mC/m^2}$ and $H$ = 3.49~nm. The gray regions indicate the boundaries of Qiao and Aluru's continuum model~\cite{Qiao05}.
}
\end{figure}


For comparison, we employ four continuum models$\colon$ a classical Gouy-Chapman (GC) model that ignores S, DE and VE effects ($a$ = $b$ = $f$ = 0), a S model ($a$ = 6.6 $\AA$ and $b$ = $f$ = 0), a DE model ($b$ =  4$\times 10^{-18}~\unit{m^2/V^2}$ and $a$ = $f$ = 0), and a VE model ($f$ =  2.3~$\times 10^{-16}~\unit{m^2/V^2}$  and $a$ = $b$ = 0). In addition, we also employ Qiao and Aluru's continuum model~\cite{Qiao05}. 
This model is basically the same as the GC model except the boundaries of eqs 10 and 11 are on the center of the first layer of water molecules and ions adjacent to walls, respectively. 
Specifically, the boundaries for $v$ and $\phi$ are $y$ = ($\pm~H$/2 $\mp$ 1.6) $\AA$ and $y$ = ($\pm~H$/2 $\mp$ 3.3) $\AA$, respectively, in this model.

Figure 2 shows profiles of $\scal[K^+]{n}$, $\scal[Cl^-]{n}$, $\phi$, $v$ and $\eta$ along the $y$-direction for five continuum models and the MD results. 
Despite the fact that it provides close estimates to the MD results in $\scal[K^+]{n}$, $\scal[Cl^-]{n}$ and $\phi$, Qiao and Aluru's continuum model greatly overestimates $v$. 
The effect of the boundary shifts on these results can be clarified by comparing the results of Qiao and Aluru's model and the GC model. 
Because the gradient of $\phi$ with respect to $y$ for the GC model is smaller than that for Qiao and Aluru's model, $v$ for the GC model is smaller than that for Qiao and Aluru's model  (still much larger than the MD results, nevertheless). Both S and DE effects do not greatly change these profiles but slightly increase $v$ especially around the channel center in comparison with the GC results. 
In contrast, when the VE effect is considered, $v$ is significantly reduced and becomes in close agreement with the MD simulation results. 
This indicates that the low $v$ is a consequence of the increased $\eta$ near the surfaces due to water molecule orientation. 
Notably, the $v$ profile obtained from the MD simulation is greatly suppressed near the surfaces. 
The increased $\eta$ near the surfaces is essential to this $v$ profile and a higher average $\eta$ cannot reproduce it.

\begin{figure} [t]
\centering
\includegraphics[keepaspectratio, width=0.46\textwidth]{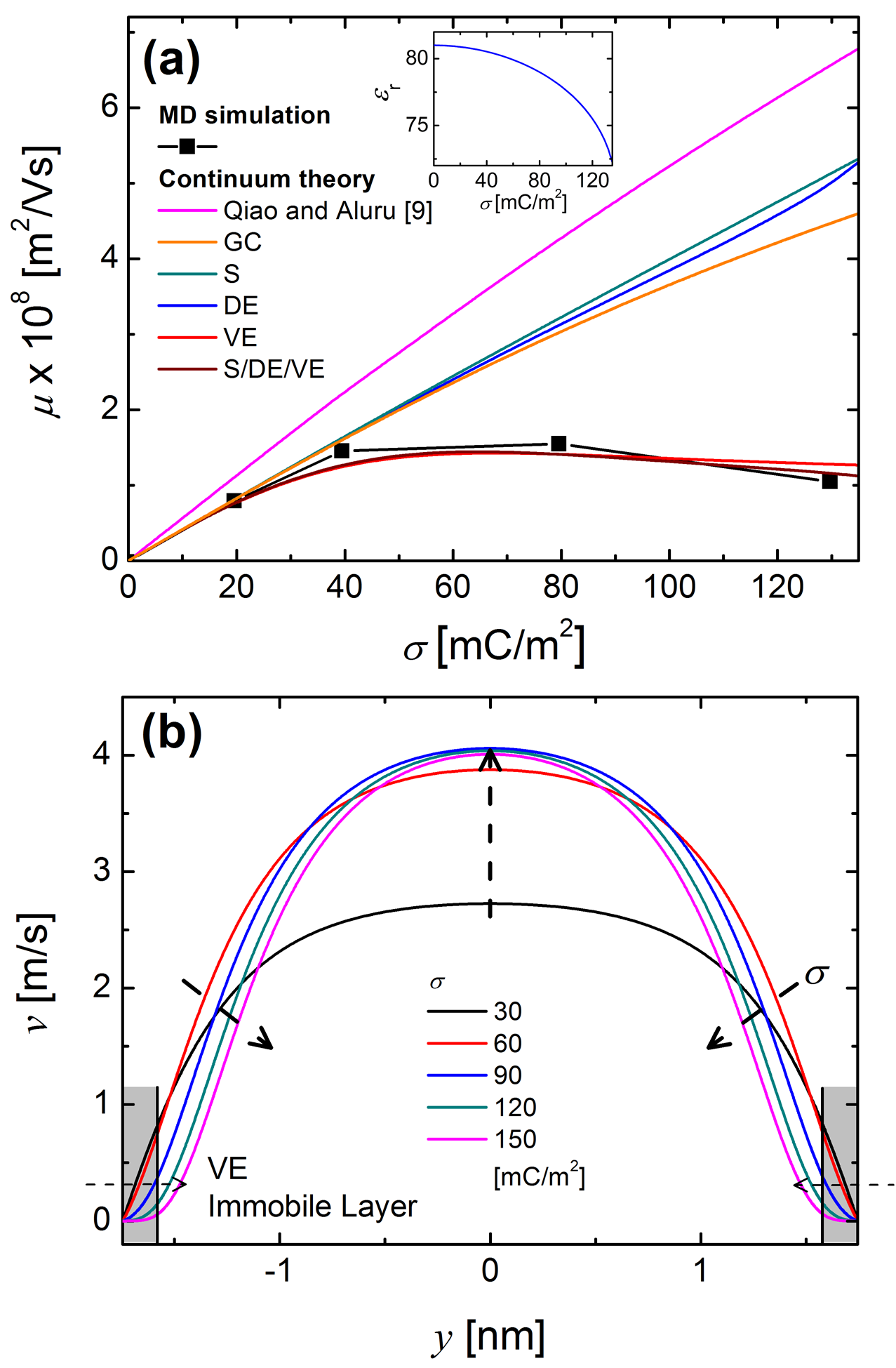}
\caption{\label{fig:wide} (a) Variation of average electroosmotic mobility $\mu$ (along with a sub-figure showing variation of relative permittivity $\scal[r]\epsilon$ at the wall interfaces as a function of $\sigma$ under the DE model) and (b) electroosmotic velocity $v$ distributions along the $y$-direction at different levels of surface charge $\sigma$  at $H$ = 3.49 nm and KCl concentration $\scal[0]{{n}}$ = 0.6 M.
}
\end{figure}

The average electroosmotic mobility $\mu$ (over the nanochannel cross sectional area) and local electroosmotic velocity at different $\sigma$ are shown in Figure 3, where $\mu$ is obtained as~\cite{Qiao05}$\colon$
\begin{equation}
\mu = \frac{1}{H}\int^{\frac{H}{2}}_{-\frac{H}{2}}\frac{v}{\vect{|E_\text{ext}|}}dy
\end{equation}
As seen in Figure~3a, the results of the VE model \emph{quantitatively} agree with the MD results over the whole range of $\sigma$. 
Importantly, $\mu$ becomes much less sensitive to $\sigma$ at high levels of  $\sigma$.
A slight decrease of $\mu$ while increasing $\sigma$ even occurs when $\sigma> 90~\unit{mC/m^2}$.
Note that, despite the hydrated ion size ($a=6.6~\AA$) being roughly 19~$\%$ of $H$ and $\scal[r]\epsilon$ near the walls dropping over 10$\%$ from its bulk value (as shown in the sub-figure of Figure 3a), it is shown that once VE effects are included, S and DE effects are significantly suppressed as evidenced by the small difference between the results from the VE model and a continuum model (S/DE/VE) that simultaneously considers S, DE and VE effects.

Figure~3b shows $v$ profiles across the nanochannel at different $\sigma$ based on the VE model.
In the low $\sigma$ regime ($i.e.$ 30 and 60 $\unit{mC/m^2}$, under which VE effects are relatively weak), the centerline electroosmotic velocity $\scal[c]{v}$ increases with the increase of $\sigma$ as does the averaged $\mu$.  
At higher $\sigma$ levels ($\geq 90~\unit{mC/m^2}$), VE immobile layers (in which the solution becomes immobile) are gradually formed in the vicinity of the walls due to the high local viscosity (highlighted in gray).

It is found that this insensitiveness of $v$ upon $\sigma$ arises from the presence of the VE immobile layers.
By double integrating eq~3 from the centerline ($dv/dy$ =0 and $v = \scal[c]{v}$) to the location at which the solution begins becoming immobile ($i.e.$ the boundary of the VE immobile layers, where $v \approx 0$), we derive a modified Smoluchowski equation$\colon$ 
\begin{equation}
\frac{\scal[c]{v}}{\vect{|E_\text{ext}|}} = \frac{\scali[0]\epsilon(\scal[r,0]\epsilon\scal[c]\phi-\scal[r,IL]\epsilon\scal[IL]\phi)}{\eta_0}
\end{equation}
where $\scal[c]\phi$, $\scal[r,IL]\epsilon$ and $\scal[IL]\phi$ are the electric potential at the centerline, relative permittivity at the boundary of the VE immobile layers and electric potential at the boundary of the VE immobile layers, respectively.
This equation indicates that when the VE immobile layers exist, $\scal[c]{v}$ is no longer determined by $\phi$ on the channel walls (which is a function of $\sigma$ by eq~10).
Instead, it depends upon a $\sigma$-independent parameter $\scal[r,IL]\epsilon\scal[IL]\phi$, rendering almost constant $\scal[c]{v}$ when $\sigma \geq 90~\unit{mC/m^2}$.
In contrast, the local $v$ (in between the centerline and the VE immobile layer boundary) decreases in response to the thickening of the VE immobile layers.
As a consequence, the derived $\mu$ decreases at larger $\sigma$.


\begin{figure} 
\centering
\includegraphics[keepaspectratio, width=0.46\textwidth]{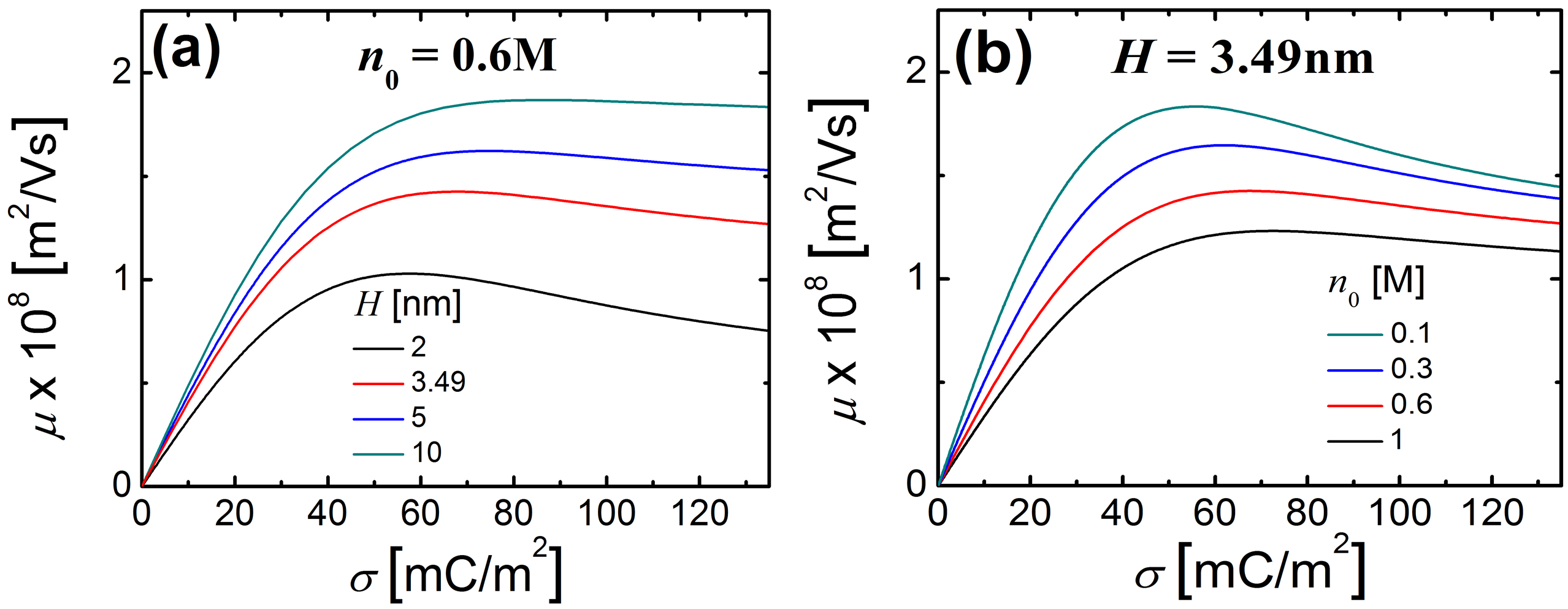}
\caption{\label{fig:wide}   Variation of average electroosmotic mobility $\mu$ as a function of surface charge density $\sigma$ (a) at different channel size $H$ and KCl concentration $\scal[0]{{n}}$ = 0.6 M, and (b) at different $\scal[0]{{n}}$ and $H$ = 3.49 nm, respectively.
}
\end{figure}

Given that the average mobility decrease is a result of the slower electroosmotic velocities within the EDLs, the degree of this decrease greatly depends on the channel size and the EDL thickness. 
At constant $\scal[0]{{n}}$ (= 0.6~M in Figure~4a), the decrease appears in the small channels ($H$ = 2 and 3.49 nm) due to the overlapped EDLs, but it almost vanishes at large $H$ (= 10 nm). 
Similarly, at fixed $H$ (=3.49~nm in Figure~4b), the decrease is apparent at the lower concentrations ($e.g.$~at $\scal[0]{n} =$ 0.1~M) when EDL overlap occurs. 
Hence, in summary, the average mobility decrease occurs under conditions where the VE immobile layer is significant at small $H$ and low $\scal[0]{n}$.

We investigate ion transport behavior using the nanochannel conductance per unit length along the channel $G$ given by$\colon$
\begin{equation}
G = \int^{\frac{H}{2}}_{-\frac{H}{2}}{\scal[L]G}dy
\end{equation}
As seen in Figure~5a, at $\scal[0]{{n}}$ = 1~M, $G$ decreases as $\sigma$ increases for the VE-modified models ($i.e.$ the VE and S/DE/VE models).
Conversely, in a lower concentration solution with $\scal[0]{{n}}$ = 0.1~M (Figure 5b), this relationship reverses at low $\sigma$, namely $G$ increases with the increase of $\sigma$, before plateauing at $\sigma \geq 60~\unit{mC/m^2}$.
These continuum based results are consistent with previous MD simulations~\cite{Qiao05}.
Note that, at high $\scal[0]{{n}}$ (= 1~M in Figure~5a), S effects, which amplify VE effects due to ion jamming~\cite{Olsson07}, become non-negligible, although the qualitative conductive behavior in response to the $\sigma$ increase remains similar.


\begin{figure} [t]
\centering
\includegraphics[keepaspectratio, width=0.46\textwidth]{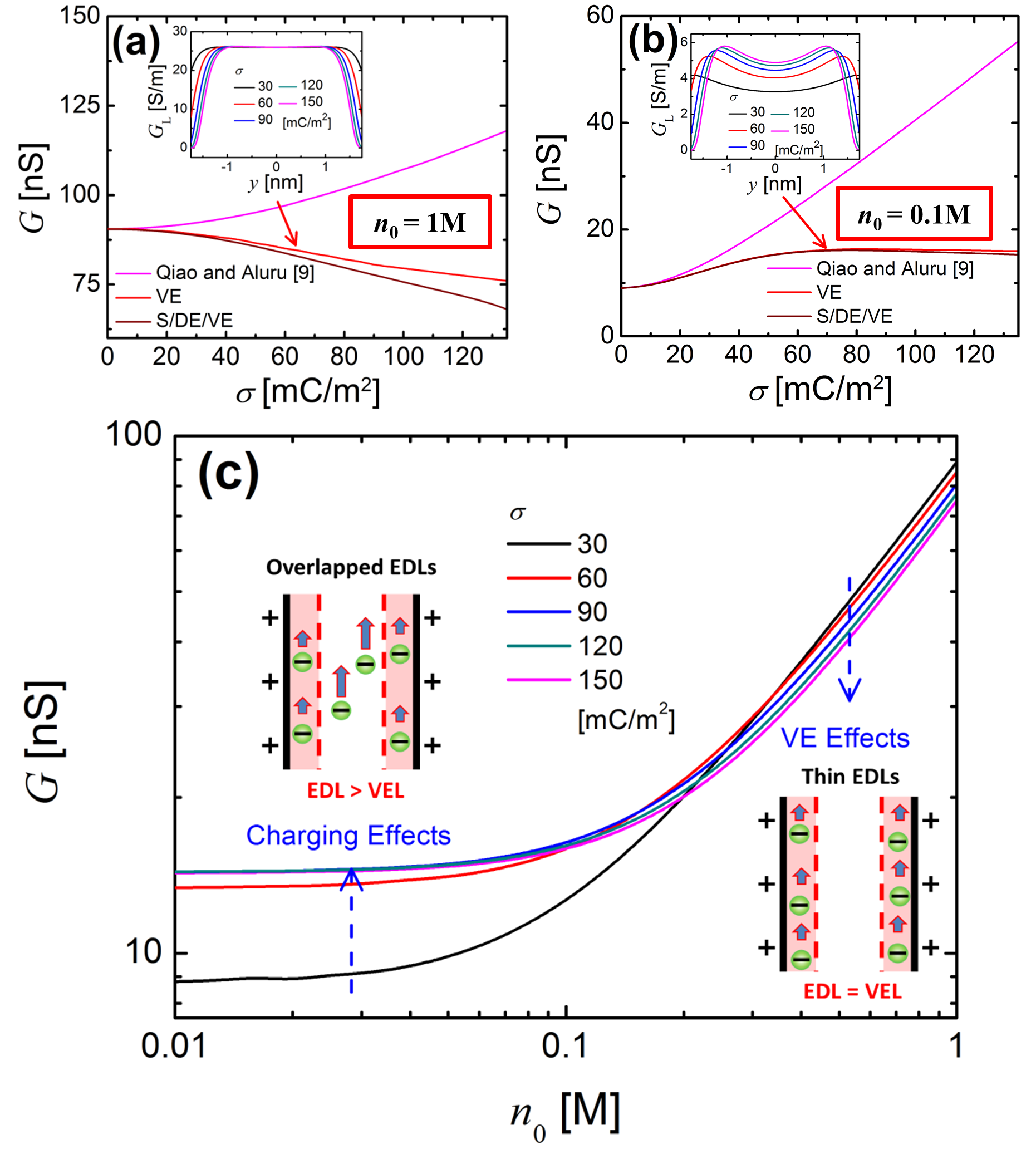}
\caption{\label{fig:wide} Variation of nanochannel conductance per unit length $G$ as a function of surface charge density $\sigma$ at $H$ = 3.49~nm and KCl concentration $\scal[0]{{n}}$ = (a) 1M and (b) 0.1M, respectively.  
Sub-figures show local nanochannel conductance per unit length $\scal[L]G$ at different surface charge density $\sigma$. 
(c) Variation of $G$ as a function of $\scal[0]{{n}}$ at different $\sigma$ levels.  
At low $\scal[0]{{n}}$ (EDL $>$ VEL) and low $\sigma$, $G$, which increases with $\sigma$, is dominated by charging effects.
At high $\scal[0]{{n}}$ (EDL $=$ VEL), VE effects dominate and $G$ decreases at larger $\sigma$.
In inset figures, the VELs are highlighted by the red shadow. 
}
\end{figure}

We herein define a viscoelectric layer (VEL), as illustrated in Figure 1,  in which VE effects are significant. 
At high concentrations, the size of the region covered by the VELs is equivalent to the EDL region. 
On the other hand, at low concentrations, when the EDLs become overlapped, the VELs only occupy the wall adjacent region. 
As a results, a `VE free zone' remains around the centerline.
At high concentrations (e.g.~$\scal[0]{{n}}$ = 1 M), as seen in the sub-figure in Figure 5a, $\scal[L]G$ near the centerline area is constant, implying that the EDLs are not overlapped. 
Within the EDLs (equivalent to VELs), in which $\scal[L]G$ is suppressed by $\sigma$, $\scal[L]G$ decreases with the increase of $\sigma$, due to higher $\eta$ and thus lower $\scali[i]{\mathcal{D}}$ (based on eq~9). 
In consequence, $G$ decreases gradually with the increase of  $\sigma$.
At low concentrations (e.g.~$\scal[0]{{n}}$ = 0.1 M), as seen in the sub-figure in Figure 5b, $\scal[L]G$ is altered with $\sigma$ across the whole range in the nanochannels. 
When $\sigma$ increases, $\scal[L]G$ is altered by two competing effects$\colon$ (i) VE effects and (ii) charging effects.
The former suppresses $\scal[L]G$ and the latter, which refers to the increase of net charge within the solution, enhances $\scal[L]G$. 
At low $\sigma$, charging effects dominate over the VE effects, while at high $\sigma$, two factors compete and offset each other.

Figure~5c shows the calculated nanochannel conductance versus KCl concentration ($G-\scal[0]{{n}}$) curves for different $\sigma$ based on the VE model. 
At high $\scal[0]{{n}}$, $G$ decreases with the increase of $\sigma$ in the same way as the previous MD simulation results~\cite{Qiao05} .

To conclude, we have evidenced that the discrepancy between the previous MD simulations and continuum theory is primarily due to the neglect of VE effects in the previous continuum model.
Two unique phenomena observed in previous MD simulations have been described by a modified continuum model that considers VE effects. 
The success of the VE-modified continuum theory shown here suggests that a similar re-examination of different electrokinetic systems at the nanoscale may also help to reduce discrepancies between results derived from continuum theory, MD simulation and experiment.

We acknowledge Prof. Rui Qiao of Virginia Tech for providing us with the MD simulation results and some additional information of ref 9.  

\bibliography{references}

\begin{thebibliography}{27}
\bibitem[{not({\natexlab{a}})}]{note1}
\bibinfo{note}{The VE coefficient $f$ was experimentally estimated to be
  $(0.5-1)~\times 10^{-15}~\unit{m^2/V^2}$ for water by Hunter and Leyendekkers
  based on eq~7,~\cite{Hunter78} which is close to Lyklema and Overbeek's
  theoretical prediction of $f= 1.02 \times
  10^{-15}~\unit{m^2/V^2}$~\cite{Lyklema61}. $f$ estimated in this study
  2.3~$\times 10^{-16}~\unit{m^2/V^2}$ is slightly lower than Hunter and
  Leyendekkers' estimate, however the difference may be due to the fact that
  $\eta$ was several times higher than $\eta_0$ near the interface in
  \citeauthor{Hunter78}'s experiments, resulting in non-negligible higher order
  field terms (which were neglected in their analysis) and a consequential
  overestimation of $f$. Other studies have similarly concluded that Lyklema
  and Overbeek's coefficient is too high~\cite{Stinger64, Hunter66}.}
\bibitem[{not({\natexlab{b}})}]{note2}
\bibinfo{note}{The MD results of $\phi$ are derived based on the MD results of
  $\scal[Cl^-]{n}$ in Figure 2a using the classical Boltzmann equation.}
\end{thebibliography}

\end{document}